# Robust Magnetic Polarons in Type-II (Zn,Mn)Te Quantum Dots

I. R. Sellers[1*#], R. Oszwałdowski[1,2], V. R. Whiteside[1], M. Eginligil[1], A. Petrou[1], I. Zutic[1], W-C. Chou[3], W. C. Fan[3], A. G. Petukhov[4] and B. D. McCombe[1]

[1] *Department of Physics, University at Buffalo SUNY, Buffalo, NY 14260 USA*

[2] *Instytut Fizyki, Uniwersytet M. Kopernika, Grudziądzka 5/7, Toruń, 87-100 PL*

[3] *Department of Electro-physics, National Chiao Tung University, Hsinchu 300, Taiwan*

[4] *Department of Physics, South Dakota School of Mines & Technology, Rapid City, SD 57701 USA*

We present evidence of magnetic ordering in type-II (Zn, Mn) Te quantum dots. This ordering is attributed to the formation of bound magnetic polarons caused by the exchange interaction between the strongly localized holes and Mn within the dots. In our photoluminescence studies, the magnetic polarons are detected at temperatures up to ~ 200 K, with a binding energy of ~ 40 meV. In addition, these dots display an unusually small Zeeman shift with applied field (2 meV at 10 T). This behavior is explained by a small and weakly temperature-dependent magnetic susceptibility due to anti-ferromagnetic coupling of the Mn spins.

2009-12-01

#current address: Sharp Laboratories of Europe Ltd, Oxford Science Park, Oxford, OX4 4GB, UK.

*electronic mail: ian.sellers@sharp.co.uk

Properties of II-VI quantum dots (QDs) enable coherent control of spin [1] and the observation of Aharonov-Bohm oscillations at elevated temperatures [2]. Doping these QDs with magnetic ions (typically Mn) [3,4,5,6,7,8] provides intriguing opportunities to manipulate and study magnetic ordering, not available in bulk magnetic semiconductors [9,10]. For example, adding an extra carrier can strongly change both the total carrier spin and the onset temperature, and [11] modifying the quantum confinement can create or destroy magnetic order, even at a fixed number of carriers [12]. Photoluminescence (PL) studies provide a sensitive probe of the magnetic properties in (II,Mn)VI QDs. For example, PL can be used as a fingerprint for the deformation of the quantum confinement and the placement of a single Mn-impurity in a QD [5]. Since the number of Mn impurities in a single QD typically does not exceed several thousand, the magnetic properties induced by carrier mediated Mn-Mn interaction differ from the situation at the thermodynamic limit. Instead, the magnetic order in QDs resembles more closely (bound) magnetic polarons (MPs) that have been extensively studied in bulk magnetic semiconductors [9,13,14,15]. During formation of MPs the exchange interaction between a localized carrier, or exciton, and magnetic impurities aligns the spins of the latter. The resulting red shift in the interband transition energy is observable in time-resolved PL [3]. Renewed interest in MP studies arises from the tunability of the QD confinement, absent for MPs formed in bulk systems.

In this work, we report a study of (Zn,Mn)Te/ZnSe QDs, characterized by the type-II band alignment, shown schematically in Fig. 1(a). The holes are strongly confined in the (Zn,Mn)Te QDs, while the electrons, forced to remain in the ZnSe matrix by the heterostructure barrier, are bound to the holes by Coulomb attraction [16]. The strong spatial separation between electrons and holes, increases the radiative lifetime by as much as two orders of magnitude [17] with respect to type-I QDs [18]. This long lifetime of the type-II excitons facilitates the formation of robust MPs, as compared to type-I magnetic QDs, where the formation is often limited by fast recombination [3]. Furthermore, in type-I magnetic QDs, for which the majority of prior experiments [3,4,5,6,7] have been performed, the interband transitions compete with the Mn internal transition at $E_{int} \sim 2.2$ eV [9] when the exciton energy, $E_x$, exceeds $E_{int}$. In the Zn(Mn,Te)/ZnSe system investigated here, this competition is absent due to the type-II band alignment, since (the indirect) $E_x < E_{int}$, in spite of both direct bandgaps being larger than $E_{int}$. To avoid the energy transfer to the Mn transitions, in some previously studied type-I systems non-magnetic dots were grown in magnetic barriers. In that geometry, however, the exchange interaction is determined mostly by the limited penetration of the hole wavefunction into the Mn-doped barriers. In Zn(Mn,Te)/ZnSe structures, the Mn ions are *inside* the QDs, resulting in a greatly enhanced overlap of holes and Mn ions, which leads to a significant enhancement in the magnetic response.

The samples were grown by molecular beam epitaxy (MBE) on a GaAs substrate. After planarization, the GaAs epilayer was transferred under high vacuum to an inter-connected II-VI MBE reactor where, following the deposition of a ZnSe buffer layer, five layers of (Zn,Mn)Te QDs were grown by migration enhanced epitaxy (MEE). The QD layers were separated by 5 nm ZnSe spacer layers, and the structure was capped with a 50 nm ZnSe layer. Cross-sectional TEM images show that the QDs form columns along the growth direction (see Fig.1b). The full details of sample growth and optimization are given in Ref. [19].

The main panel in Fig. 1 shows $\sigma^{+,-}$ (positive, negative helicity) components of the PL of the (Zn,Mn)Te/ZnSe QDs at temperature $T$ = 4.2 K and zero applied magnetic field, $B$. The PL is excited non-resonantly by the linearly polarized 2.41 eV line of an Ar$^+$ laser above the bulk (Zn,Mn)Te band-gap (~2.4 eV). Both $\sigma^{+,-}$ components display a PL peak at ~ 1.9 eV, arising from exciton recombination across the type-II indirect gap of the dots. This transition is well below the energy of the Mn internal transition (~2.2 eV), and consequently the deleterious effects of the Mn $d$ states, discussed above, are avoided.

While (II,Mn)VI QDs are expected to be paramagnetic with $\sigma^{+,-}$ PL components coinciding at $B$ = 0, surprisingly we find a difference between the corresponding intensities $I^{+,-}$, suggesting that the spin-degeneracy of the lowest levels of electrons or holes (or both species) is lifted. Remarkably, the resulting circular polarization $P = (I^+ - I^-)/(I^+ + I^-)$ persists up to at least $T$ ~ 165 K [20], which may be consistent with the presence of holes in some of the QDs in the absence of light, as expounded below, and in independent magnetization measurements, which provide evidence (hysteresis) for an ordered magnetic state in the dark [21].

Prompted by this unexpected behavior, we have performed continuous wave (CW) and time-resolved PL measurements. Both (II,Mn)VI bulk and QD structures are known to display a giant excitonic Zeeman splitting $\Delta$ (50-100 meV) in applied magnetic fields. The splitting $\Delta$ is characterized by the magnitude of the effective g-factors (at low fields where $\Delta$ vs. field is linear, and a g-factor can be defined), which can exceed 100 [8,9,10], but rapidly decreases with $T$. With the strong magneto-optical coupling in (II,Mn)VI QDs, the $B$- and $T$-dependences of PL can thus provide further information about the ordering of the carrier and Mn spins. The trends in the dependence of the circular polarization on $B$, shown in Fig. 1(c) for the Faraday geometry ($B$ applied along the direction of propagation of light), are typical for paramagnetic (II,Mn)VI QDs, and are consistent with $\Delta$ increasing with $B$ and decreasing with $T$. However, unexpectedly, $\Delta$ is more than an order of magnitude smaller than found for paramagnetic bulk samples of (Zn,Mn)Te [9] (for Mn content between 5 and 11%, typical values in QDs fabricated by the same growth procedures [19,22]). At $B \sim 10$ T the shift in the peak of the total PL intensity, which is roughly equal to $\Delta/2$ at high $B$, is only $\sim 2$ meV [16].

To explore the possibility that this unusual effect could be attributed to the formation of magnetic polarons, which frequently appear in type-I magnetic QDs, we have investigated the temperature dependence of time-resolved (TR) PL at $B = 0$. The PL decay reveals a long recombination lifetime $\tau_R \sim 4$ ns, which allows for uninterrupted MP formation in QDs. Figure 2(a) shows the normalized TR PL spectra in increments of 800 ps. Initially, the PL peak energy is $\sim 1.93$ eV, reaching an equilibrium value of $\sim 1.90$ eV after a few ns. As in the previous measurements on type-I QDs, we attribute this evolution of the PL peak to the formation of MPs [3]. In this process, the hole's spin aligns the randomly oriented Mn spins which, in turn, produce Zeeman splitting [13] of the hole levels. The PL peak position at $t = 0$ corresponds to the exciton recombination energy prior to alignment of the Mn spins. During the subsequent alignment, the hole occupies preferentially the lower Zeeman level and therefore the exciton PL peak undergoes a red shift with a characteristic MP formation time $\tau_{MP}$.

Figure 2(b) shows the peak energy of the PL as a function of time at various temperatures. The MP formation is distinguishable up to $\sim 200$ K, an order of magnitude higher $T$ than in (Zn,Mn)Se/CdSe QDs [3] and even higher than observed for MPs in (Cd,Mn)Te QDs (120 K) [4] excited resonantly with circularly polarized light [23]. In fact, using higher excitation powers, we find evidence of MPs up to 270 K. At the higher powers, however, dipole (band bending) effects at the type-II interface could (at least partially) obscure the MP red-shift complicating analysis. [24]. Thus we discuss only results obtained at low power excitation, which constrained the temperature range for reliable results to $T < 150$ K, since the PL signal is quenched due to ionization of the excitons at higher $T$. Additionally, we have used a non-magnetic ZnTe/ZnSe QD sample grown by the same procedure, to separate the dipole effects, which we find only significantly affect the PL energy determination at times greater than 10 ns.

To determine the MP binding energy $E_{MP}$, and the formation time $\tau_{MP}$ at different $T$, we fit the temporal evolution of the PL peak by a double-exponential decay, rather than the single exponent, used by others [25,3,26]. We find two separate timescales for all $T$; on average $\tau_1 \sim 0.7$ ns, and $\tau_2 \sim 11$ ns. We attribute $\tau_1$ to the process of alignment of Mn moments with the spin of a hole. Magnetic polaron formation with two very different timescales has also been observed recently for colloidal QDs [27]. The slower process was attributed to directional re-orientation of the MP due to crystal anisotropy, but is unlikely to be the *only* origin of $\tau_2$ in these strongly anisotropic QDs. Although the exact nature of the longer lived regime is not well understood, it is likely that the inhomogeneity of the QDs *also* contributes to the unusual temporal behavior. Additional work is now underway to elucidate further the underlying physics of this complex system. Using the fits, we find a total $E_{MP} = E_1 + E_2 \sim 40\pm0.9$ meV with a surprisingly weak *T*-dependence, Fig. 2(c), and $\tau_{MP} = \tau_1 + \tau_2 \sim 12$ ns.

While our CW and TR measurements appear consistent with the formation of MPs, there are noticeable differences with the previous studies in magnetic QDs. Our unusual findings, such as the weak *T*-dependence of $E_{MP}$ and the persistence of MPs at high temperatures, warrant further discussion. Comparison of the PL peak positions in the CW and TR spectra, reveals that the former correspond to fully formed MPs, which is consistent with $\tau_{MP} < \tau_R$ [28]. Thus, it may be surprising that *P* is greatly reduced at 80 K (see Fig. 1c), even though $E_{MP}$ is much larger than the corresponding thermal energy.

To explain this apparent paradox, we have developed a simple model of MP formation based on models used for QWs [29]. First, we neglect the influence of electrons on the magneto-optical properties, since both their exchange interaction and wavefunction overlap with Mn spins are much smaller than those of holes. We assume that at short times after its creation, the hole occupies a spin-up or a spin-down level in a QD, with thermodynamic occupancy probabilities given by the initial spin splitting of hole levels, $\Delta_i^h$, determined by the *external* magnetic field *B* and by *T*, see Fig. 3(a). The preferred orientation of the hole spin is along the growth direction (the quantization axis). The process of aligning the Mn spins via the hole-Mn exchange takes place on a significantly longer timescale. Throughout this process, the probability of a spin-flip to the higher hole level decreases, as the level splitting, $\Delta^h(t)$, increases to reach its final value of $2E_{MP} \approx 80$ meV, determined now by the large *internal* molecular field $B_{\text{int}}$. Thus, it is predominantly the initial $\Delta_i^h$ that determines the proportion of MPs with hole spin parallel versus anti-parallel to *B* (At B = 0, MPs are fully formed after a few nanoseconds, but there is an equal population of spin-up and spin-down.). Furthermore, $\Delta_i^h$ does not exceed $k_B \times 80$K even at 10 T. In the framework of this physical picture, a built-in magnetic field existing *before* photo-creation of holes would lead to a finite $P$ at $B = 0$. This built-in field could exist in a subset of QDs containing *equilibrium* holes [6]. Magnetization measurements provide evidence for an aligned spin-state in the absence of photo-injected holes [21]. An alternative explanation, such as remnant magnetization of a spin-glass phase of Mn [30], is less likely due to the exceptionally high temperatures to which the $P$ persists.

The measurements of $E_{\mathrm{MP}}(T)$ in our QDs [Fig. 2(c)] reveal a very weak $T$-dependence, which is strikingly different from that of previously studied II-VI DMS QDs [3]. We believe that this behavior reflects an anomalous $T$-dependence of the magnetic susceptibility of the Mn spin system, $\chi(T)$. To study $\chi(T)$, we first note that the magnetic response of the QD system is most likely in the linear regime at 150K. Since we find $E_{MP}$ and the zero-field circular polarization to be essentially $T$-independent up to 150K, we assume a linear response of the magnetization even at the lowest experimental temperatures. In this regime, as is the case for donor-bound MPs [13], the two observable quantities $E_{MP}$ and $\chi(T)$ can be related to each other by (S.I. units)

$$E_{MP} = \mu_0^{-1} \left( J_{ex} / 2g\mu_B N_0 \right)^2 \eta \left( E_{\mathrm{MP}} / k_B T \right) \Omega_{e\!f\!f}^{-1} \chi(T) \quad , \tag{1}$$

where $J_{ex}$ is the exchange integral for holes, $N_0$ is the cation density, $g = 2$, $\mu_B$ is the Bohr magneton, and $\Omega_{e\!f\!f}$ is the effective volume of the MP [31], defined by $\Omega_{e\!f\!f}^{-1} = \int |\psi(\mathbf{r})|^4 d\mathbf{r}$. The term $\eta\left(E_{MP}/k_B T\right)$ interpolates between the two limiting cases of strong, $\eta(x) = \tanh(x)$, and weak, $\eta(x) = \tanh x + 2/x$, magnetic anisotropy [13]. The former case is essentially the Ising model of a MP [14], while the latter case corresponds to fully isotropic spin fluctuations [13].

First we use Eq. (1) to investigate the magnitude of $E_{MP}$ at low $T$ (where $\eta = 1$ for both models). The very small PL peak shift with $B$, [16], indicates an exceptionally low $\chi$. We attribute the low susceptibility to a combination of antiferromagnetic (AFM) Mn coupling and shape anisotropy of the QDs. The strong AFM character of Mn-Mn interaction is consistent with bulk [32] and epitaxial [33] MnTe properties. Since Mn acts as a nucleation center for the growth of the QDs in MEE, its concentration will be considerably higher than 5-11% *in the dots* [34], enhancing the AFM behavior.

The low susceptibility must be compensated by a small $\Omega_{eff}$ (strong confinement) to obtain the large $E_{MP}$. For anisotropic 3D parabolic confinement for a disk-shaped QD with diameter $d$ and height $h$, one finds for the ground state $\Omega_{eff} = (\pi\hbar)^{3/2}(2m^*E_b)^{-3/4}d\sqrt{h}$, where $E_b$ is the (Zn,Mn)Te/ZnSe valence band offset. With $E_b = 1$ eV, [35], $d = 20$ nm, $h = 3$ nm, and $m^* = 0.5m_0$, this expression gives $\Omega_{eff}$ an order of magnitude smaller than the QD volume. Further reduction of $\Omega_{eff}$ may result from additional hole localization during MP formation. Similar strong hole localization effects were also found in type-II magnetic quantum wells [36].

The value of $E_{MP}$ found in our experiments is almost 3 times larger than in Ref. [3], even though the p-d exchange integrals in (Zn,Mn)Te ($J_{ex}$=1.05eV) and (Zn,Mn)Se ($J_{ex}$=1.11eV) are very similar [9]. We attribute this difference to enhancement of exchange-interaction due to incorporation of Mn in the QDs, rather than in the barriers, as explained above.

The relatively large error bars for $E_{MP}(T)$ at higher temperatures result from our use of very small excitation powers to avoid possible band bending. To study the unusual $T$ dependence of $\chi$ without the experimental scatter, below we use a linearized $E_{MP}(T)$,

Fig. 2(c). We consider first the vanishing anisotropy case, and substitute $E_{MP}(T)$ into Eq. (1) and calculate $\chi(T)$. The results are shown in Fig. 3(b) by the solid squares. The system behaves like a "paramagnet" with a strong AFM character. The temperature decrease is much weaker than for the conventional type-I magnetic quantum dots, as extrapolated from the low-$T$ behavior of $E_{MP}(T)$ given in Ref. [3].

The fully isotropic case is improbable, but it is important to consider it, as it provides a lower bound for $\chi(T)$. We believe that our system is much closer to the high anisotropy case; $\eta(x) = \tanh(x)$, with the easy axis perpendicular to the growth direction. Two anisotropy factors must be taken into account: 1) that of the magnetic susceptibility, $\chi_{//} \neq \chi_{\perp}$, where $\chi_{//}$ is the component in the growth direction, and 2) that of the heavy hole $g$-factor, i.e. $g_{//} \gg g_{\perp}$, due to strong spin-orbit coupling and a quasi-2D shape of the hole wave function. The second factor causes the system to behave as an Ising-like MP with $T$-dependence following that of $\chi_{//}(T)$, shown in Fig. 3(b) by the open circles. We propose the following scenario: in the absence of holes the Mn spins in the QDs are in-plane, coupled anti-ferromagnetically. After photo-excitation, the nonequilibrium holes form MPs with magnetization in the z-direction caused by canting of the Mn spins, which acquire a small out-of-plane component. The overall weak $T$-dependence is therefore consistent with the transverse susceptibility of antiferromagnets [37]. The smallness of $\chi_{\perp}$ also explains the relatively long spin alignment time, $\tau_1 \sim 700$ ps, as compared to $\tau_{MP} <$ 200 ps [3].

In conclusion, we describe measurements that reveal formation of exciton magnetic polarons that persist at unusually high temperatures in type-II magnetic QDs.

This intriguing system combines a weak response to the external magnetic field with a strong Mn-hole exchange coupling. We expect that our findings will stimulate future experimental studies of type-II magnetic QDs aimed at tailoring of room temperature magnetism in these structures. In particular, gate-control of magnetic order and the possibility of interacting magnetic polarons could offer versatile opportunities for using these magnetic dots in spin-based information processing [11,12,38].

This work is supported by the ONR, NSF-ECCS, NSF-ECCS CAREER, AFOSR, NSC 98-2119-M-009-015 the CCR and the Office of the Provost at the University at Buffalo, SUNY, and the CNMS ORNL.

**Figure 1**

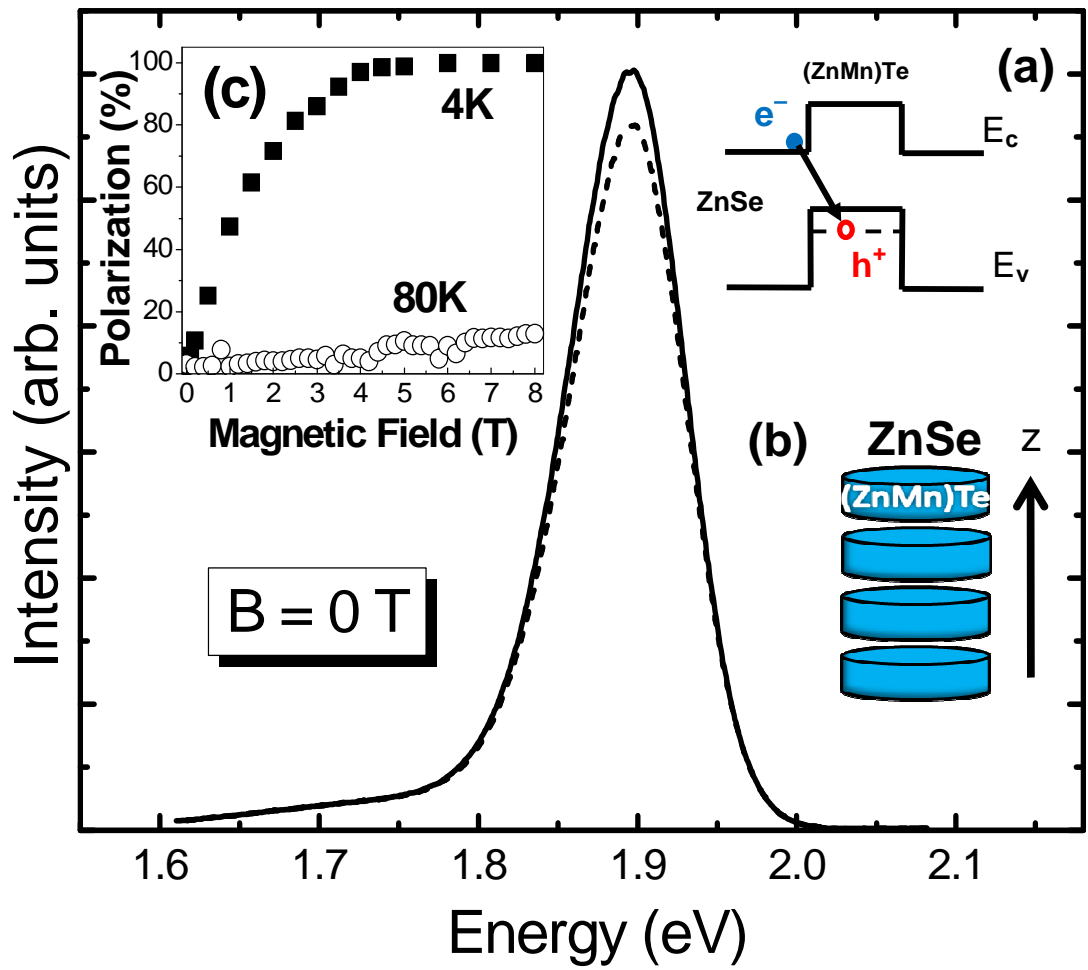

Figure2

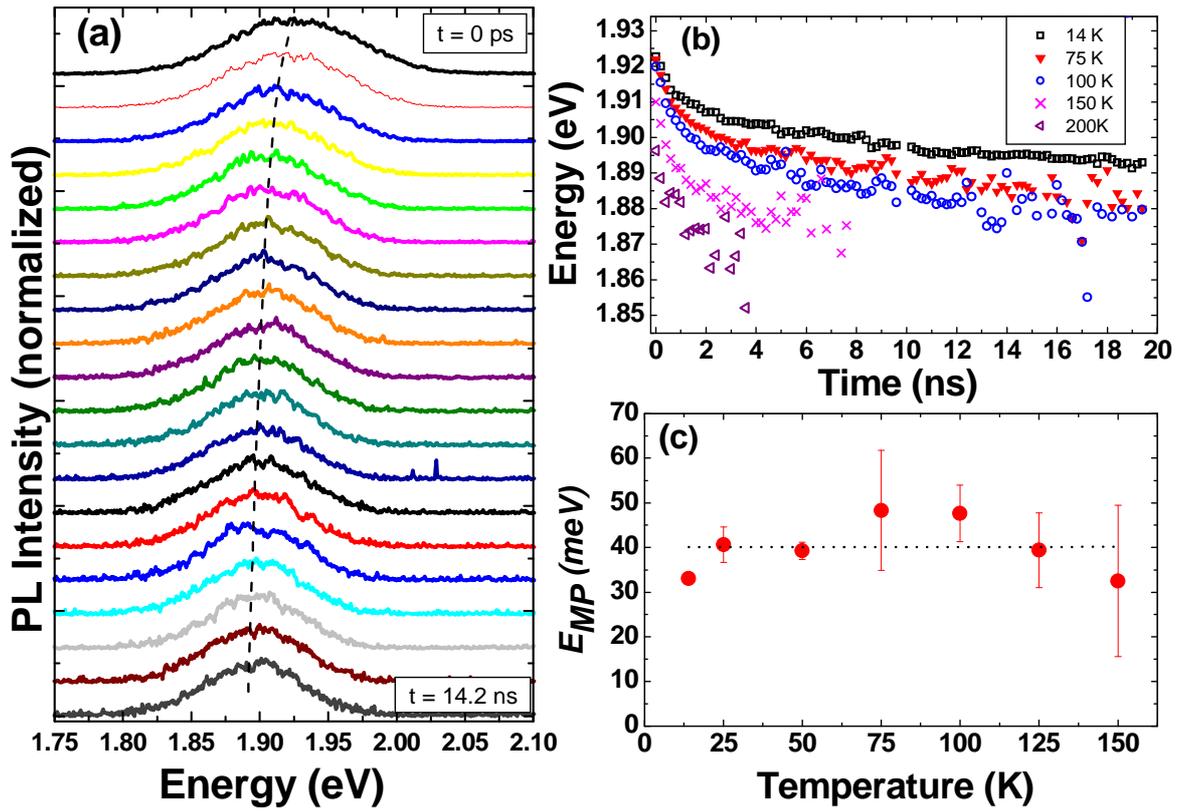

**Figure 3**

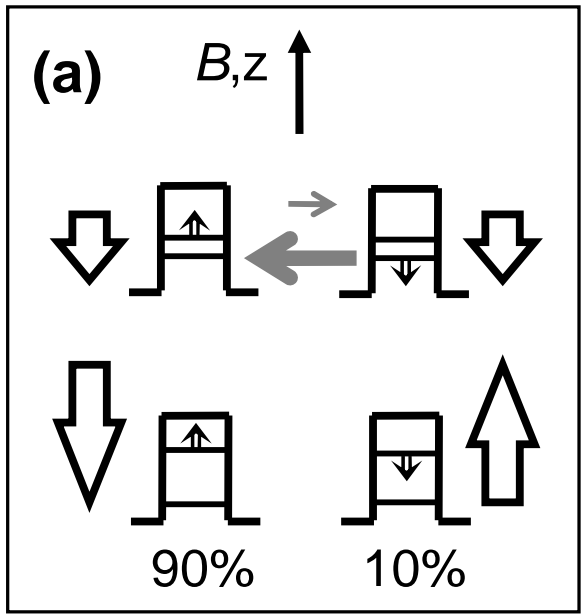
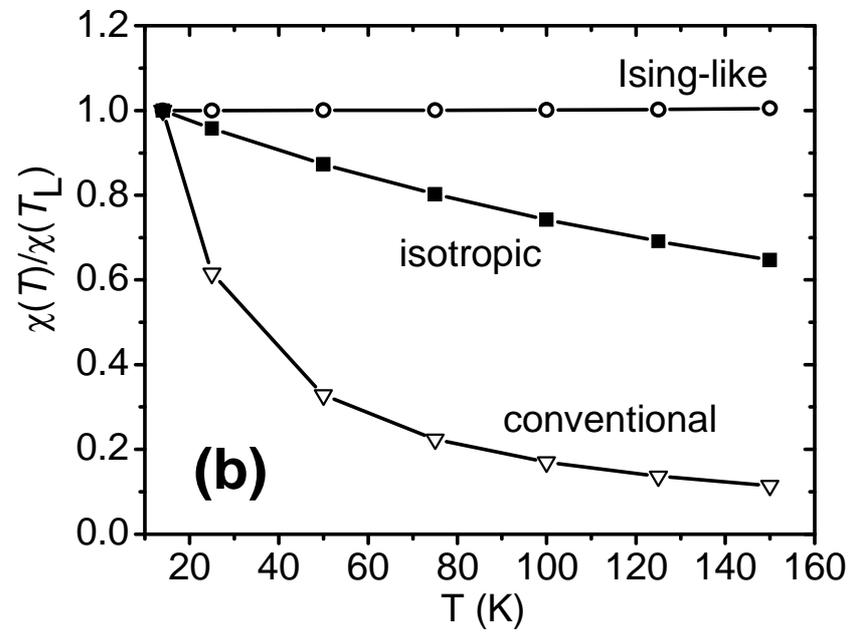

**Figure Captions**

**Figure 1.** Zero magnetic field (*B*) photoluminescence at 4.2 K analyzed for positive ($\sigma^+$, solid line) and negative ($\sigma^-$, dashed line) helicity. Continuous wave linearly polarized excitation is at 2.41 eV. Inset **(a)** shows type-II band alignment and the hole confinement. Inset **(b)** shows the columnar geometry of the quantum dots. **(c)** Magnetic field (*B*) dependence of the circular polarization of the photoluminescence (PL) at 4.2 K (squares) and 80 K (circles).

**Figure 2. (a)** Temporal evolution of the PL at 14 K in 800ps. The PL peak intensities are normalized to maximum intensity at time t=0. **(b)** Time dependence of the peak PL energy for temperatures ranging from 14 to 200 K, supporting the formation of magnetic polarons (MPs). The overall energy shift of the different curves reflects the thermal reduction of the band gap. **(c)** Temperature dependence of the MP binding energy ($E_{MP}$) with a linear fit (black line).

**Figure 3. (a)** Schematic diagram illustrating magnetic polaron (MP) formation in external magnetic field *B*. Arrows: (full-line) hole spins; (hollow) z-components of Mn spins; (horizontal grey) relative magnitudes of spin-flip rates driving the hole occupancies towards equilibrium. *Top*: initial stage where the hole-level splitting is defined by the external *B*. *Bottom*: final stage. Numbers give a possible percentage of QDs that will host spin-up and -down MPs. **(b)** Temperature dependence of susceptibility $\chi$ determined from $E_{MP}(T)$ in Fig. 2(c) and different QD models: fully isotropic magnetic fluctuations (squares), and highly anisotropic, Ising-like system (circles). For comparison we show $\chi$ in conventional magnetic QDs from Ref. 3 (triangles). All $\chi$'s are normalized to their values at the lowest experimental temperature, $T_L$=14K.